\PassOptionsToPackage{table,xcdraw}{xcolor}
\documentclass[sigconf]{acmart}
\usepackage[table]{xcolor}
\makeatletter
\newcommand\subsubsubsection{\@startsection{paragraph}{4}{\z@}{-2.5ex\@plus -1ex \@minus -.25ex}{1.25ex \@plus .25ex}{\normalfont\normalsize\bfseries}}
\newcommand\subsubsubsubsection{\@startsection{subparagraph}{5}{\z@}{-2.5ex\@plus -1ex \@minus -.25ex}{1.25ex \@plus .25ex}{\normalfont\normalsize\bfseries}}
\makeatother

\acmConference[CASCON'23]{33rd Annual International Conference on Computer Science and Software Engineering}{Sep. 11-14 2023}{Las Vegas, USA}

\usepackage[outercaption]{sidecap}    
\usepackage[figuresright]{rotating}
\usepackage{booktabs}
\usepackage{soul}
\usepackage{subfigure}
\usepackage{pgfplots}
\usepackage{pgfplotstable}
\usepackage[export]{adjustbox}
\pgfplotsset{width=5.5cm}
\usepackage[framemethod=TikZ]{mdframed} 

\AtBeginDocument{%
  \providecommand\BibTeX{{%
    \normalfont B\kern-0.5em{\scshape i\kern-0.25em b}\kern-0.8em\TeX}}}

\usepackage{balance}  
\usepackage{graphics} 
\usepackage{color}
\usepackage{booktabs}
\usepackage{ccicons}

\usepackage{url}
\usepackage{fancybox}
\usepackage{multirow}
\usepackage{flushend}
\usepackage{booktabs}
\usepackage{tabularx}
\usepackage{comment}
\usepackage{array}
\usepackage[flushleft]{threeparttable}
\usepackage{mdframed}
\graphicspath{{}{images/}{dia/}}
\DeclareGraphicsExtensions{.pdf,.png,.svg}

\usepackage{listings}
\usepackage{courier}
\usepackage{hyperref}
\usepackage{xpatch}

\xpatchcmd{\refstepcounter}{%
  \stepcounter{#1}%
}{%
  \stepcounter{#1}%
  \xdef\scr{\number\value{#1}}%
}{\typeout{success}}{\typeout{failure}}

\newcounter{o}
\usepackage{tikz}
\definecolor{1c1}{RGB}{188,162,6}
\definecolor{1c2}{RGB}{137,129,80}
\definecolor{1c3}{RGB}{239,167,31}
\definecolor{1c4}{RGB}{88,194,241}
\definecolor{1c5}{RGB}{6,180,188}

\usepackage{listings}

\tikzset{mynode/.style={draw=white,solid,circle,fill=green,inner sep=1pt, thick,
text=black}}
\tikzset{arrow line/.style={dashed, line width= 2.5pt, color=#1}}
\usepackage{varwidth}
\def\bf{\textbf}

\def\fig {Figure~}

\def\tbl {Table~}

\def\sec {Section~}

\usepackage{paralist}

\newcommand{\nd}{\vspace{1mm}\noindent}
\usepackage{tikz}

\lstdefinestyle{inlinecode}{basicstyle={\ttfamily\scriptsize\bfseries}}

\newcommand{\urls}[1]{{\scriptsize\url{#1}}}
\usepackage{tcolorbox}

\usepackage{paralist}
\usepackage[outercaption]{sidecap}
\usepackage{caption}
\usepackage [autostyle, english = american]{csquotes}
\MakeOuterQuote{"}
\newcounter{scn}
\usepackage[shortlabels]{enumitem}
\usepackage{tikz}
\usetikzlibrary{trees,positioning,shapes,shadows,arrows}
\usepackage[edges]{forest}
\usepackage{pgf-pie}
\usepackage{paralist}
\usepackage{soul}
\usepackage[utf8]{inputenc}
\usepackage[english]{babel}
\usepackage{float}
\usetikzlibrary{positioning}

\usetikzlibrary{positioning,shadows}
\usepackage{balance}

\newif\ifpienumberinlegend
\pgfkeys{/number in legend/.code=
    \expandafter\let\expandafter\ifpienumberinlegend
    \csname if#1\endcsname
    \ifpienumberinlegend

    \def\beforenumber##1\afternumber{}%
    \fi,
    /number in legend/.default=true
}
\definecolor{1c1}{RGB}{188,162,6}
\definecolor{1c2}{RGB}{137,129,80}
\definecolor{1c3}{RGB}{239,167,31}
\definecolor{1c4}{RGB}{88,194,241}
\definecolor{1c5}{RGB}{6,180,188}

\usepackage{tcolorbox}
\tcbuselibrary{listings,skins}
\tikzset{mynode/.style={draw=white,solid,circle,fill=green,inner sep=1pt, thick,
text=black}}
\tikzset{arrow line/.style={dashed, line width= 2.5pt, color=#1}}

\usepackage{bchart}
\usepackage{xcolor}

\tikzstyle{chart}=
[legend label/.style={font={\small},anchor=west,align=left},
legend box/.style={rectangle, draw, minimum size=5pt},
axis/.style={black,thin,->},
axis label/.style={anchor=east,font={\tiny}}]

\tikzstyle{bar chart}=[
chart, bar width/.code={
    \pgfmathparse{##1/2}
    \global\let\bar@w\pgfmathresult},
bar/.style={thin, draw=black},
bar label/.style={font={\bf\small},anchor=north},
bar value/.style={font={\footnotesize}},
bar width=.75,]

\tikzstyle{pie chart}=
[chart,
slice/.style={line cap=round, line join=round, thin, draw=black},
pie title/.style={font={}},
slice type/.style 2 args={
    ##1/.style={fill=##2},
    values of ##1/.style={}}]

\pgfdeclarelayer{background}
\pgfdeclarelayer{foreground}
\pgfsetlayers{background,main,foreground}

\newcommand{\tempsum}{0}

\newcommand{\pye}[3][]{
    \begin{scope}[#1]
        \pgfmathsetmacro{\curA}{0}
        \pgfmathsetmacro{\r}{1}
        \def\c{(0,0)}
        \node[pie title] at (270:1.3) {#2};
        \def\tempsum{0}
        \foreach \v/\s in{#3}{
          \pgfmathparse{\v+\tempsum}
          \global\let\tempsum=\pgfmathresult}
        \foreach \v/\s in{#3}{
            \pgfmathsetmacro{\deltaA}{\v/\tempsum*360}
            \pgfmathsetmacro{\nextA}{\curA + \deltaA}
            \pgfmathsetmacro{\midA}{(\curA+\nextA)/2}
            \path[slice,\s] \c
            -- +(\curA:\r)
            arc (\curA:\nextA:\r)
            -- cycle;
            \pgfmathsetmacro{\d}{min(1.2,max((1.5-0.1*\v) , .5)}

            \begin{pgfonlayer}{foreground}
                \path \c -- node[pos=\d,pie values,values of \s]{$\v\%$} +  (\midA:\r);
            \end{pgfonlayer}
            \global\let\curA\nextA}
    \end{scope}}

\def\test#1{%
    \ifnum #1 > 0
      #1
    \fi
}

\newcommand{\lstbg}[3][0pt]{{\fboxsep#1\colorbox{#2}{\strut #3}}}
\lstdefinelanguage{diff}{
  morecomment=[f][\lstbg{red!20}]-,
  morecomment=[f][\lstbg{green!20}]+,
  morecomment=[f][\textit]{@@},
}

\newcommand{\definebox}[2]{%
  \newcounter{#1}
  \newenvironment{#1}[1][]{%
    \stepcounter{#1}%
    \mdfsetup{%
        frametitle={%
            \tikz[baseline=(current bounding box.east),outer sep=0pt]
            \node[anchor=east,rectangle,fill=white]
            {\strut \MakeUppercase#1~\csname the#1\endcsname\ifstrempty{##1}{}{:~##1}};}}%
    \mdfsetup{innertopmargin=0pt,linecolor=#2,%
        linewidth=0.5pt,topline=true,
        frametitleaboveskip=\dimexpr-\ht\strutbox\relax,}%
    \begin{mdframed}[]\relax%
    }{\end{mdframed}}%
}

\definebox{Recommendation}{black}

\usepackage{hyperref}
\usepackage{color}
\usepackage{balance}
\usepackage{tcolorbox}
\begin{document}

\title{Characterizing Issue Management in Runtime Systems}

\settopmatter{authorsperrow=3}
\author{Salma Begum Tamanna}
\affiliation{
    \institution{DISA Lab, University of Calgary} 
    \country{Calgary, Alberta, Canada}
 } 
 \email{salmabegum.tamanna@ucalgary.ca}

\author{Gias Uddin}
 
\affiliation{
    \institution{DISA Lab, University of Calgary} 
      \country{Calgary, Alberta, Canada}
 }
\email{gias.uddin@ucalgary.ca}
\author{Lan Xia and Longyu Zhang}

\affiliation{
    \country{IBM, Ottawa, Ontario, Canada}
}
\email{lan_xia@ca.ibm.com}
\email{longyu.zhang@ibm.com}

\balance

\begin{abstract}
Modern programming languages like Java require runtime systems to support the implementation and deployment of software applications in diverse computing platforms and operating systems. 
These runtime systems are normally developed in GitHub-hosted repositories based on close collaboration between large software companies (e.g., IBM, Microsoft) and OSS developers. However, despite their popularity and broad usage; to the best of our knowledge,  these repositories have never been studied. We report an empirical study of around 118K issues from 34 runtime system repos in GitHub. We found that issues regarding enhancement, test failure and bug are mostly posted on runtime system repositories and solution related discussion are mostly present on issue discussion. 82.69\% issues in the runtime system repositories have been resolved and 0.69\% issues are ignored; median of issue close rate, ignore rate and addressing time in these repositories are 76.1\%, 2.2\% and 58 days respectively. 82.65\% issues are tagged with labels while only 28.30\% issues have designated assignees and 90.65\% issues contain at least one comment; also presence of these features in an issue report can affect issue closure. Based on the findings, we offer six recommendations to ensure effective management of the runtime system issues.

\end{abstract}




\maketitle
\section{Introduction} \label{sec:intro}
Most programming languages require the support of a framework, referred to as runtime system that provides a platform on which programs can be executed. A runtime system may interface with low level functions like  processor or memory management and translation of the high level program syntax to binary code; also with high level functions which include the services of code generation, debugging and optimization or type checking to prevent execution of flawed code. For a number of years now, numerous runtime systems have been open-sourced on platforms like GitHub in order to leverage the distributed development process and developers or users can raise issues there about any flaw they find on the program or ask question or request enhancement.

We have found several attempts are taken where researchers empirically investigated various GitHub communities from AI to blockchain \cite{stateofai,blockchain,AIrepo,covid} to identify unique properties, pattern and trends in those ecosystems. However, despite their popularity and broad usage; to the best of our knowledge,  these
repositories have never been examined to identify  diverse states of the runtime system issue reports or issue discussion contents along with the issue management patterns and factors influencing issue closure.  Intuitively, such insights can help us better support the runtime system ecosystems.  

We report an empirical study of around 118K issues from 34 runtime system GitHub repositories. We analyzed how the issues are reported and discussed and how the issues are managed. Based on the analysis, we answer five research questions.

\noindent\textbf{RQ1. What types of issues are found in the runtime system repositories?}
We extracted statistically significant sample number of issues (383) from all 34 repositories and manually labelled the issues with the reason they are reported by checking issue description and created a taxonomy of issues for the runtime system reports. We found that runtime system issue reports can be of 10 types (e.g, bug, enhancement, build issue, etc.). 

\noindent\textbf{RQ2. How are the issues addressed?}
82.69\% issues in the runtime system repositories are closed where 30.04\% issues are closed by the issue poster. Among the open issues,  4.00\% are found to be ignored. Median of issue close rate, ignore rate and addressing time is 76.1\%, 2.2\% and 58 days respectively.

\noindent\textbf{RQ3. How are the issues managed?}
Issues are frequently associated with user-defined tags, referred to as `labels' by the developers. 82.65\% of issues are tagged with labels while only 28.30\% issues have designated assignees and 90.65\% issues contain at least one comment.

\noindent\textbf{RQ4.  What are the relationships between metadata used to describe the issues and the closure of issues?}
Presence of comments, labels, and designated assignees and number of contributors in a repository can have a statistically significant correlation with the closure of an issue.

\noindent\textbf{RQ5. Is there any relationship between the types of information discussed in issue comments and the closure of the issue?}
13 different types of information can be discussed in the issue comments like solution discussion, social conversation, etc. None of them have any statistically significant correlation with issue closure.

We offer six recommendations based on our study findings that can be used by runtime system developers and maintainers to devise policies and guidelines to better manage runtime system issues.

\section{Background} \label{sec:back}
\subsection{Runtime System}
Unlike assembly language where programmers need to write low level instructions for specific platforms, programming languages like Java, Python, JavaScript etc. offer higher level of abstraction with the support of polymorphism, generics, concurrency, automatic memory management and other advanced features. To execute program on different platforms regardless of the programming language being used \cite{languageruntime}, an intermediate platform independent support which will abstract the instructions for any platform and hardware, becomes a necessity. To facilitate the implementation and execution of these languages; beyond a static compiler, a high level language virtual machine or runtime system was introduced. For M programming languages and N target platforms, this platform independent intermediate representation is able to reduce the number of translation from M*N to M+N; thus facilitating the implementation of programming language across diverse hardware architectures.  

Runtime systems are used extensively for years and evolved to provide many more functionalities. Typically  runtime systems dynamically translate high level program syntax to binary code for target platforms. Apart from this, these systems involve in processor or memory management by setting up or managing stack and heap or by garbage collecting.  Monitoring code and execution like accessing variables, passing parameters between procedures, interfacing with operating system and debugging can be other features. Also, sometimes a runtime system is able to manage code for performance improvement or type check to prevent execution of flawed code or generate code.

\subsection{Runtime System Issues}
 GitHub hosted runtime system repositories exploit the issue driven development by ensuring that besides project maintainers, testers and developers; outsiders like users of open source runtime systems can also report any kind of issues. Typically stakeholders of runtime systems report on any bug or error they found while building, compiling, running or testing the system. Issues are also used to ask questions, request for enhancement or simply start a discussion. \fig{\ref{fig:issue} shows an example of a runtime system issue report  where the issue creator opens this issue to point out the need of documentation. An issue report typically contains a title as highlighted in the figure, that summarizes the issue in short; followed by the issue body which describes the issue elaborately. Besides text, code snippets, images and URLs can also be added to the issue body. An issue can be in two states: Open and Closed. When an issue is reported, it remains open and after resolving, maintainers can close it to mark the resolution. For issue management and addressing, managers have the ability to distribute issues to the responsible experts or contributors on his own can engage to resolve issues. An issue can also be tagged with labels to mark the types of issue and stakeholders can  comment on issues for discussing with interested parties. 
  \begin{figure}[t]
\hspace{-7pt}
\includegraphics[scale=0.14]{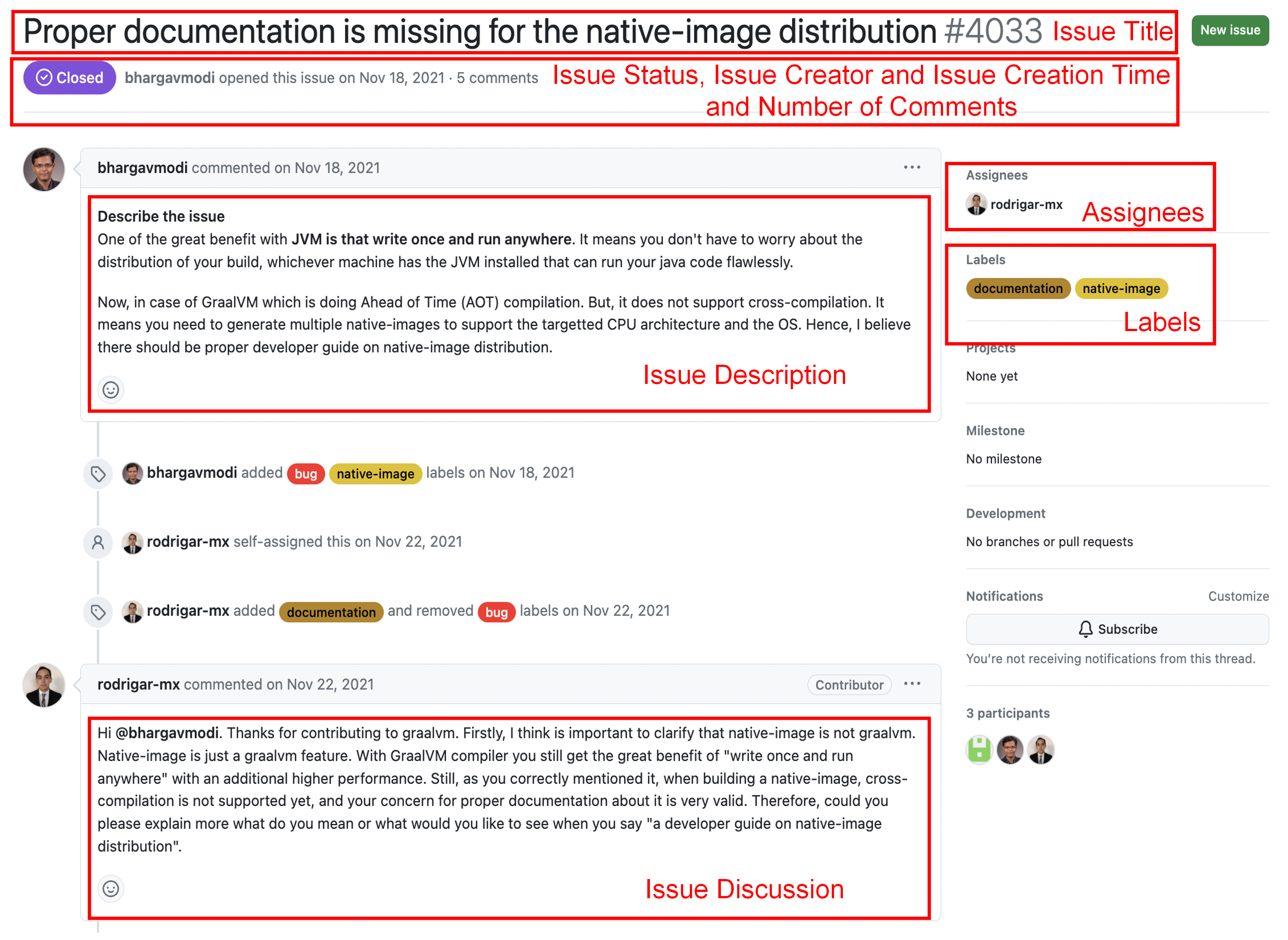}
\caption{A runtime system GitHub issue report}
\label{fig:issue}
\end{figure}

\section{Studied Data} \label{sec:data}
This section explains the data collection process and characteristics of the selected runtime system repositories for our empirical study. 

\subsection{Project Selection} 

We used GitHub platform to search for open source runtime system repositories for our empirical study. GitHub serves more than 372 million repositories from 100 million developers as of January 2023, making it the biggest source code server to date. A number of professionally and independently created runtime systems are also hosted on GitHub.  Popular runtime systems like Mono, NodeJs and OpenJ9 are hosting their repositories on GitHub from 2010, 2014 and 2017 respectively. In order to get a list of the runtime system repositories, we used GitHub search bars to look for repositories matching with certain keywords like \textit{runtime}, \textit{runtime system}, \textit{jvm}, \textit{virtual machine} etc. Similarly, we also made use of the GitHub tag to search for related repositories. Using these techniques, we could obtain 50 repositories. However, when we further checked whether these repositories are actually runtime systems or not; we need to discard some wrong selection. As our next step, we discarded the ones with no issues since in our empirical study, we aimed to look at the issues to understand the characteristics unique to the runtime systems. We also removed the repositories on which issues are written in non-English language. Finally 34 (68\%) runtime system repositories remained that we used in our study.
\vspace{1mm}

\subsection{Data Retrieval} 

We retrieved the repository metadata and issue data from these chosen repositories using the GitHub API. Repository metadata includes number of stars/forks/watchers, list of contributors and labels, creation/last modification date and number of open and closed issues. Issue information includes retrieving issue title, body and discussion along with its status, creation and closing date (if any) and list of assignees or labels.  We also listed a statistically significant samples of issues with 95\% confidence level and 5\% confidence interval for a part of our study. We selected those 383 issues at random from all of the repositories in a way that ensured at least one issue was chosen from each source. 

\subsection{Data Characteristics} 

All the 34 repositories combined contain total of 117789 issues where 46.2\% (54435) issues are of \textit{.Net} project and 13.5\% (15844) of \textit{NodeJs}. All the other projects combined contain rest (40.3\%) of the issues. \tbl\ref{table:projects} presents the project list with number of open and closed issues, stars, forks and contributors. Based on the number of stars and forks, NodeJs and Deno are the topmost popular repositories of all.

\renewcommand{\tabcolsep}{3pt}
\begin{table}[t]
\rowcolors{2}{}{lightgray!30}
    \centering
    \caption{Studied Projects (P) sorted by \# issues. \#O/\#C = \# open/closed issues, \#Star = \# stars \#Ct = \# contributors} 
    
    \begin{tabular}{l|p{23mm}|l|l|l|l|l|l}
    \toprule
        P & Name & \#O & \#C & Year & \#Star & \#Fork & \#Ct\\    \midrule
        1 & dotnet/runtime & 8249 & 46186 & 2019-23 & 11721 & 3867 & 313 \\ 
        2 & nodejs/node & 1292 & 14552 & 2014-23 & 94708 & 25690 & 443 \\ 
        3 & denoland/deno & 1220 & 6459 & 2018-23 & 88946 & 4821 & 436 \\ 
        4 & eclipse-openj9/openj9 & 1991 & 4902 & 2017-23 & 3078 & 661 & 197 \\ 
        5 & mono/mono & 2115 & 2819 & 2010-23 & 10384 & 3777 & 219 \\ 
        6 & microsoft/\newline onnxruntime & 1417 & 3050 & 2018-23 & 8877 & 2057 & 378 \\ 
        7 & oracle/graal & 831 & 2661 & 2016-23 & 18462 & 1492 & 246 \\ 
        8 & dapr/dapr & 350 & 2466 & 2019-23 & 20931 & 1645 & 211 \\ 
        9 & erlang/otp & 262 & 2057 & 2009-23 & 10400 & 2895 & 346 \\ 
        10 & chakra-core/\newline ChakraCore & 515 & 1605 & 2016-23 & 8850 & 1316 & 147 \\ 
        11 & containerd/\newline containerd & 365 & 1626 & 2015-23 & 13707 & 2809 & 439 \\ 
        12 & tokio-rs/tokio & 230 & 1666 & 2016-23 & 20162 & 1884 & 446 \\ 
        13 & bytecodealliance/\newline wasmtime & 507 & 1276 & 2017-23 & 11992 & 970 & 367 \\ 
        14 & jerryscript-project/\newline jerryscript & 102 & 1382 & 2015-23 & 6530 & 660 & 95 \\ 
        15 & cri-o/cri-o & 76 & 1112 & 2016-23 & 4505 & 935 & 245 \\ 
        16 & wasmerio/wasmer & 212 & 933 & 2018-23 & 14963 & 637 & 141 \\ 
        17 & LuaJIT/LuaJIT & 49 & 667 & 2015-23 & 3754 & 814 & 0 \\ 
        18 & bytecodealliance/\newline wasm-micro-runtime & 143 & 481 & 2019-23 & 3676 & 470 & 115 \\ 
        19 & containers/crun & 15 & 354 & 2017-23 & 2151 & 238 & 77 \\ 
        20 & containers/youki & 45 & 177 & 2021-23 & 4605 & 260 & 76 \\ 
        21 & awslabs/aws-lambda-rust-runtime & 5 & 214 & 2018-23 & 2674 & 266 & 73 \\ 
        22 & ikvm-revived/ikvm & 69 & 139 & 2018-23 & 707 & 84 & 15 \\ 
        23 & apigee/trireme & 34 & 108 & 2012-23 & 468 & 53 & 10 \\ 
        24 & ReadyTalk/avian & 35 & 106 & 2012-23 & 1208 & 174 & 47 \\ 
        25 & microsoft/napajs & 59 & 63 & 2017-23 & 9240 & 347 & 32 \\ 
        26 & smol-rs/smol & 13 & 103 & 2020-23 & 2782 & 133 & 34 \\ 
        27 & runtimejs/\newline runtime & 40 & 72 & 2014-23 & 1913 & 132 & 5 \\ 
        28 & cloudflare/\newline workerd & 45 & 51 & 2022-23 & 4570 & 159 & 46 \\ 
        29 & lunatic-solutions/lunatic & 33 & 39 & 2020-23 & 4055 & 121 & 30 \\ 
        30 & losfair/blueboat & 18 & 25 & 2021-23 & 1876 & 54 & 2 \\ 
        31 & elsaland/elsa & 18 & 21 & 2020-23 & 2757 & 58 & 16 \\ 
        32 & pojala/electrino & 18 & 7 & 2017-23 & 4314 & 105 & 7 \\ 
        33 & YaroslavGaponov/\newline node-jvm & 8 & 16 & 2013-23 & 2095 & 182 & 6 \\ 
        34 & voodooattack/\newline nexusjs & 5 & 8 & 2016-23 & 1081 & 22 & 3 \\

        \bottomrule
    \end{tabular}
    \label{table:projects}
\end{table}

\subsubsection{Result} \tbl\ref{table:relationship} presents the relationship of 11 features with issue closure by showing the values of the Cohen’s $\delta$ and effect size \ along with the statistical significance from Wilcoxon rank-sum test \cite{Wilcoxon1992}. This result shows that all the features show a statistically significant difference between the open and closed issues except sentiment of title and body. Again, the effect size of has-label is small (0.23); meaning labelled issues are more likely to be closed. Similarly, has-assignee, num-assignee and has-comment has a small effect size of 0.27, 0.24  and 0.3 respectively which indicates that issues are more likely to be closed if these have assignees or there are multiple assignees or there are discussion under the issues. Also, num-contributors feature also shows a small effect size of 0.24 indicating  the more contributors in a repository, the more issue will be closed. The other remaining features shows negligible effect size. 

\section{Empirical Study}\label{sec:empiricalstudy}
We answer five research questions by analyzing the issues reported in the studied runtime system repos: 
\begin{enumerate}[label=\textbf{RQ\arabic{*}.}]
    \item What types of issues are found in the runtime system repositories?
    \item How are the issues addressed?
    \item How are the issues managed?
    \item What are the relationships between metadata used to describe the issues and the closure of issues?
    \item Is there any relationship between the types of information discussed in issue comments and the closure of the issue?
\end{enumerate}
In RQ1, we aimed at analyzing issue report contents to understand the issue types raised by the stakeholders. In RQ2 and RQ3, we shared the current states on how runtime system issues are being addressed and managed. For this, we have measured the issue close rate, ignore rate and average addressing time for open and closed issues and investigated the presence of labels and assignees on the issue reports and repositories. Finally we checked what factors are influencing the issue closure in this ecosystem. RQ4 presents the effect of different issue or repository metadata on issue closure. RQ5 discusses the relationship between closure of the issue and information type found in issue discussion .
\begin{table*}[t]
\rowcolors{2}{}{lightgray!30}
    \centering
        \caption{The distribution of sample issues in different categories}
    \label{table:category}
    \begin{tabular}{l|l|l}
    \toprule
        \textbf{Category} & \textbf{Definition} & \textbf{Total} \\ \midrule
        Enhancement & Request of new features, functions, support or services or any modification & 91 (23.76\%) \\
        Test Failure & Test failures from Jenkins nightly or weekly build or other sources &  61 (15.93\%)\\
        Bug & Any abnormal behaviour or wrong, invalid and unexpected output of the program &  59 (15.40\%) \\
        Question & Any function usage query, query on upcoming features and request for explanation or confirmation & 40 (10.44\%)\\
        Runtime Error & Any exception or error while running the program, crash or segmentation fault & 39 (10.18\%)\\
        Build Issue & Build failures, installation issues or issues occured after upgradation  & 39 (10.18\%)\\
        Memory Issue &  Memory leak, heap overflow and need of memory & 19 (4.96\%) \\
        Performance Issue & Slow process, long compilation time or running time &  10 (2.61\%)\\
        Documentation Issue & Missing or wrong documentation &  9 (2.35\%)\\
        Others & Other issues like compilation error, managing tasks for automation or participation request & 16 (4.18\%)\\
        \bottomrule
    \end{tabular}
\end{table*}
\subsection{RQ1. What types of issues are found in the runtime system repositories?} \label{sec:issuetype}
%
\subsubsection{Approach} We considered a statistically significant number of sample issues for the manual categorization of issue reports. Selected 383 sample issues are randomly picked from all 34 repositories maintaining the fact that at least one issue report has been picked from each repository. First author manually labelled the issues with single category by investigating issue title and description. We used the open card sorting method for categorizing the issues where we had no predefined categories; rather we created new category for labelling. The labelling tells the message why an issue is raised rather than the root cause. We then grouped some categories if we think these can be fallen under same category. Finally we obtained the taxonomy of the runtime system issues. 

\subsubsection{Result} Based on our manual analysis of the sample issues, we obtain total ten broad categories that indicate why an issue has been raised. \tbl\ref{table:category} presents the definition of each category and the number (and percentage) of the issues labelled under this category. 

Our analysis shows that 23.76\% (91) of the sample issues are about Enhancement request; making it the largest category. Users often request for new features or modification to the runtime system project maintainers. Also, volunteers, developers or managers are often found to raise Enhancement issues explaining possible implementation process. As an example of this category, we can consider \#2181 issue of P12 where users ask to the team to provide a macro explaining the situation where this can be useful (See \fig\ref{fig:issuecategory}).  The second and third mostly found issue types are Test failure (15.93\%) and Bug (15.40\%). \#32882 issue of P01 is an example of Test Failure report sharing details and \#4978 of P10 is of a Bug where user raised an issue to report the wrong treatment of long negative number.  We distinguish bug reports from enhancement ones by defining that enhancement reports are for requesting changes for added support; not due to any bugs of the program. Total 40, 39 and 39 issues are found in the category of Question, Runtime Error and Build Issue respectively. \fig\ref{fig:issuecategory} shows an example of a question (\#1853)  that has been asked by a user to P03 members to know either this project support a specific format of code. Memory, Performance and  Documentation Issue category each covers below 5\% of the total issues.  \fig\ref{fig:issuecategory} holds an example of the mentioned first two categories and \fig\ref{fig:issue} illustrates a issue report on missing documentation. However, we did not categorize 16 issues to any particular type as they were low in number as a group and thus we put them into a single category as Others.

    \begin{figure*}[t]
    \centering
    \includegraphics[scale=0.25]{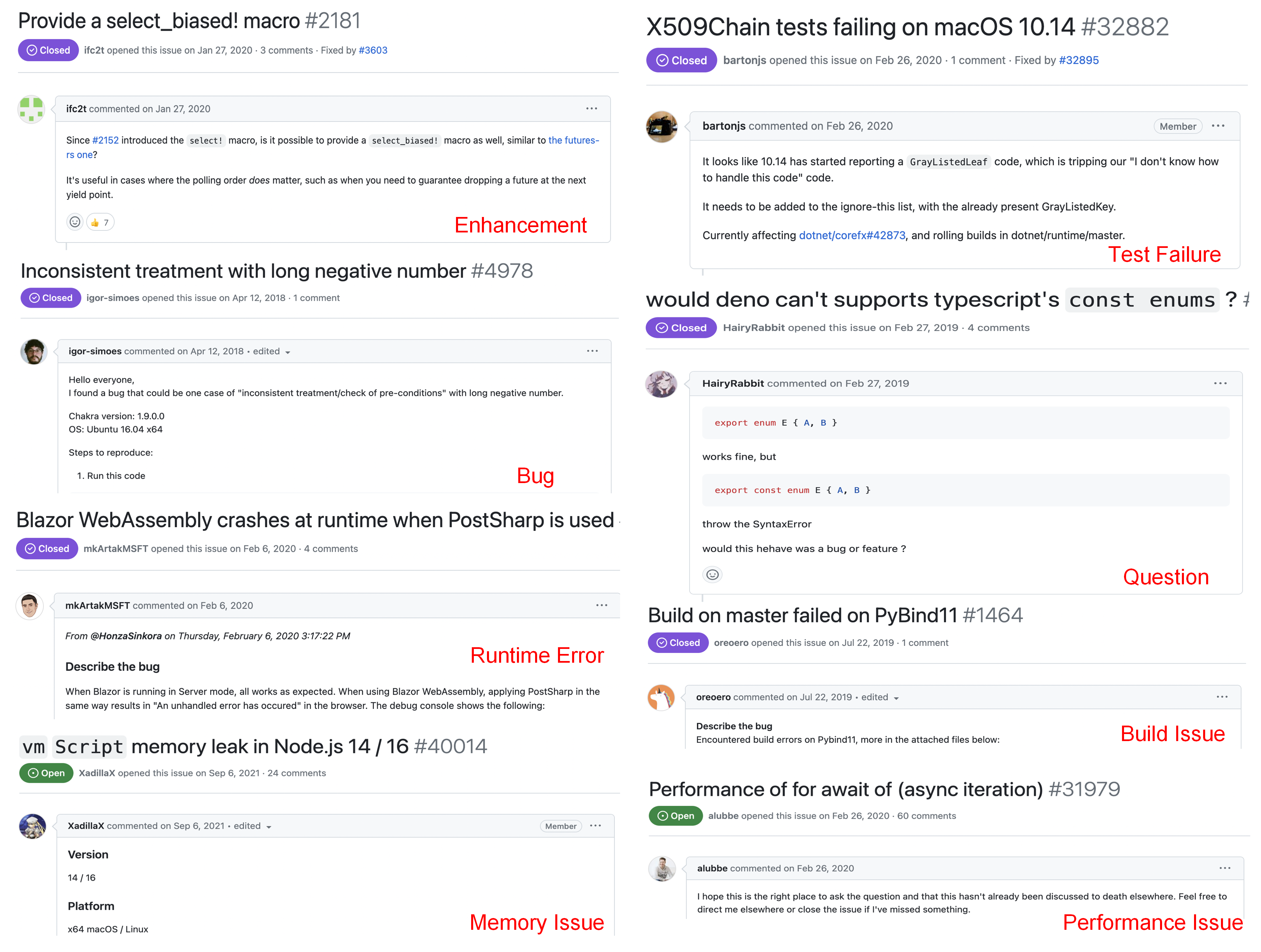}
    \caption{Example of different types of issue reports}
    \label{fig:issuecategory}
    \end{figure*}

\begin{tcolorbox}[flushleft upper, boxrule=1pt, arc=0pt, left=0pt, right=0pt, top=0pt, bottom=0pt, colback=white, after=\ignorespacesafterend\par\noindent]
\textbf{Summary of RQ1.} We took 383 sample issues from the selected runtime system repositories and manually categorized them based on the reason stakeholders raise these issues. We obtained total ten categories of issue reports. Enhancement reports are the mostly found issues; followed by Test Failure and Bug.
\end{tcolorbox}

\subsection{RQ2. How are the issues addressed?}\label{sec:issueaddress}
    
   \subsubsection{Approach} First using GitHub API, we extracted the issue status of all issues of the selected projects. Issue status indicates whether an issue is resolved or not. Unresolved issues are categorized as open issues and resolved ones as closed issues. We calculated the number of open and closed issues; then dividing them by the total number of issues, we obtain issue open/close rate for each repository. In similar manner, we have also calculated the issue ignore rate for each. Ignored issues are those which does not receive any response like comment, developer assignment or labelling from issue maintainers after opening the issues. Next, we extracted the  creation date and closed date (if any) of the issues and measured the addressing time per closed issue to know the time it remained open. We further investigated the information about issue creator and resolver to find out how many issues are self closed; meaning the issue creators themselves closed these issue after resolution.
   \begin{figure}[t]    
    \centering
    \hspace{-3px}
        {\includegraphics[scale=0.47]{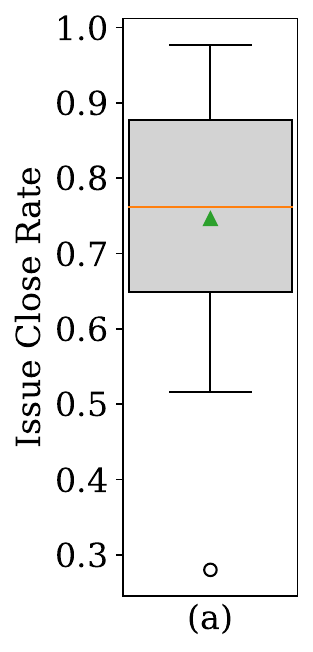}}   
        \hspace{1px}
        {\includegraphics[scale=0.47]{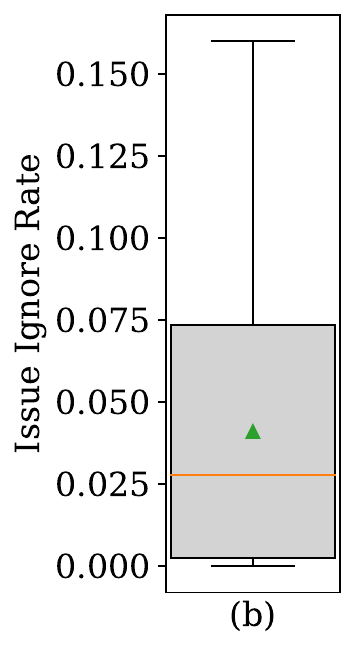}}
        \hspace{1px}
        {\includegraphics[scale=0.47]{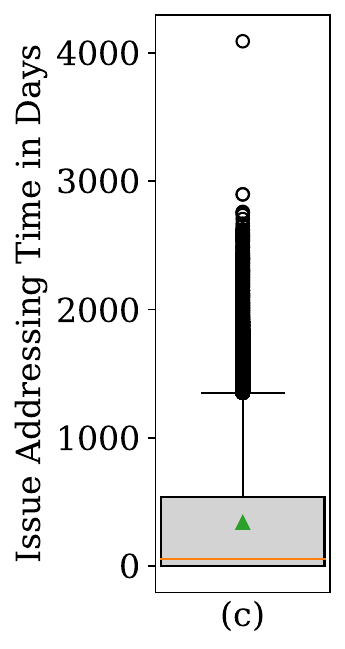}}
        \caption{The distribution of (a) issue close rate, (b) issue ignore rate and (c) issue addressing time in the selected runtime system repositories.}
        \label{fig:distribution}
    \end{figure}
       
    \subsubsection{Result} Only 17.31\% (20386) issues are remained open while the remaining 82.69\% (97403) issues are resolved in selected runtime system repositories. \fig\ref{fig:distribution}(a) shows the distribution of issue close rate per repository. We find that except one (P32), all the other repositories have close rate greater than 50\% and the average and median issue close rate are around 75\% and maximum is 97.72\%. Also among these closed issues, 30.0\% (29224) issues are self-closed and among open issues, only 4.00\% (815) issues are ignored ones. \fig\ref{fig:distribution}(b) illustrates the distribution of issue ignore rate where mean is 4\% and median is 2.2\%. Only 2 repositories have issue ignore rate greater than 12\% and 4 projects have no ignored issues; specially .Net instead of the having the largest amount of issues has issue ignore rate of zero. Finally, in  \fig\ref{fig:distribution}(c), the distribution of issue addressing time is shown. It can be observed that the average addressing time of issues, standard deviation and longest addressing time are 339.69, 494.92 and 11.2 years respectively for the closed issues. However, the median is only 58 days which reflects even if we ignore the anomalies, still issue addressing takes a huge amount of time; can go up to two months. 

\begin{tcolorbox}[flushleft upper, boxrule=1pt, arc=0pt, left=0pt, right=0pt, top=0pt, bottom=0pt, colback=white, after=\ignorespacesafterend\par\noindent]
\textbf{Summary of RQ2.} We investigated the distribution of issue close rate, ignore rate and addressing time across selected repositories to understand how issues are addressed in the runtime system projects. The result shows 82.69\% issues are closed of all issues, 30.04\% issues are self closed of all closed issues and  4.00\% are ignored of all open open issues. Median of issue close rate, ignore rate and addressing time is  76.1\%, 2.2\% and 58 days.
\end{tcolorbox}

\subsection{RQ3. How are the issues managed?}\label{sec:issuemanagement}
    
    
   \subsubsection{Approach} To analyze the practices the of runtime system issue management, we extracted the list of issue labels, assignees  and number of comments per issue for all issues of the selected projects. Next we calculated the number of issues with labels, assignees and comments of every repository and find the distribution of comments.  We also investigated the unique labels found on runtime system repositories and listed down the the topmost used labels.
   \subsubsection{Result} The study result shows almost all the runtime system repositories of our list which is 94.12\% (32) use labels for issue categorization and 82.65\% (97351) issues of all are also tagged with issue labels. Among the closed issues, 81.27\% (79158) have labels and among the open issues,  89.24\% have labels. All of these 32 repositories introduce new labels besides the default ones provided by GitHub. Including the default ones, there can be found total 1154 unique labels. .Net and NodeJS being the largest two repositories used more labels than others which is 286 and 195 respectively. On the other hand, 13 repositories contain labels equal or more than 50 and 5 of the repositories have less than 10 labels.  \fig\ref{fig:labels} shows the 10 mostly found labels on the runtime system repository where bug being the highest; followed by question. Since .Net has the most issues, some of the unique labels of this repository also falls into this list. This finding of mostly used labels also support our analysis of issue types of the sample issues in RQ1 since frequency of the issues related to bug, question, enhancement and test failure are also found more than others in the sample issues.

\begin{figure}[t]
    \centering
    \begin{tikzpicture}  
    \begin{axis}  
    [axis x line*=bottom,
     axis y line*=left,
     height = 54mm,
    xbar,  
    xlabel={\ Count}, 
    bar width=4pt,
    symbolic y coords={ area-System.Net,test failure,tenet-performance,os-linux,help wanted,untriaged,enhancement,area-CodeGen-coreclr,question,bug},
    ytick=data,  
    nodes near coords, 
    nodes near coords align={horizontal},  
    ]  
\addplot [fill=gray,
        ]coordinates { (2179,area-System.Net) (2280,test failure)
(2729,tenet-performance) (2814,os-linux) (3269,help wanted) (3871,untriaged) (4218,enhancement) (5135,area-CodeGen-coreclr) (5353,question) (12745,bug)
};  
\end{axis} 
\end{tikzpicture}  
\caption{Most frequently used 10 labels in runtime system repositories}
\label{fig:labels}
\end{figure}
\begin{figure}[t]
\includegraphics[scale=0.3]{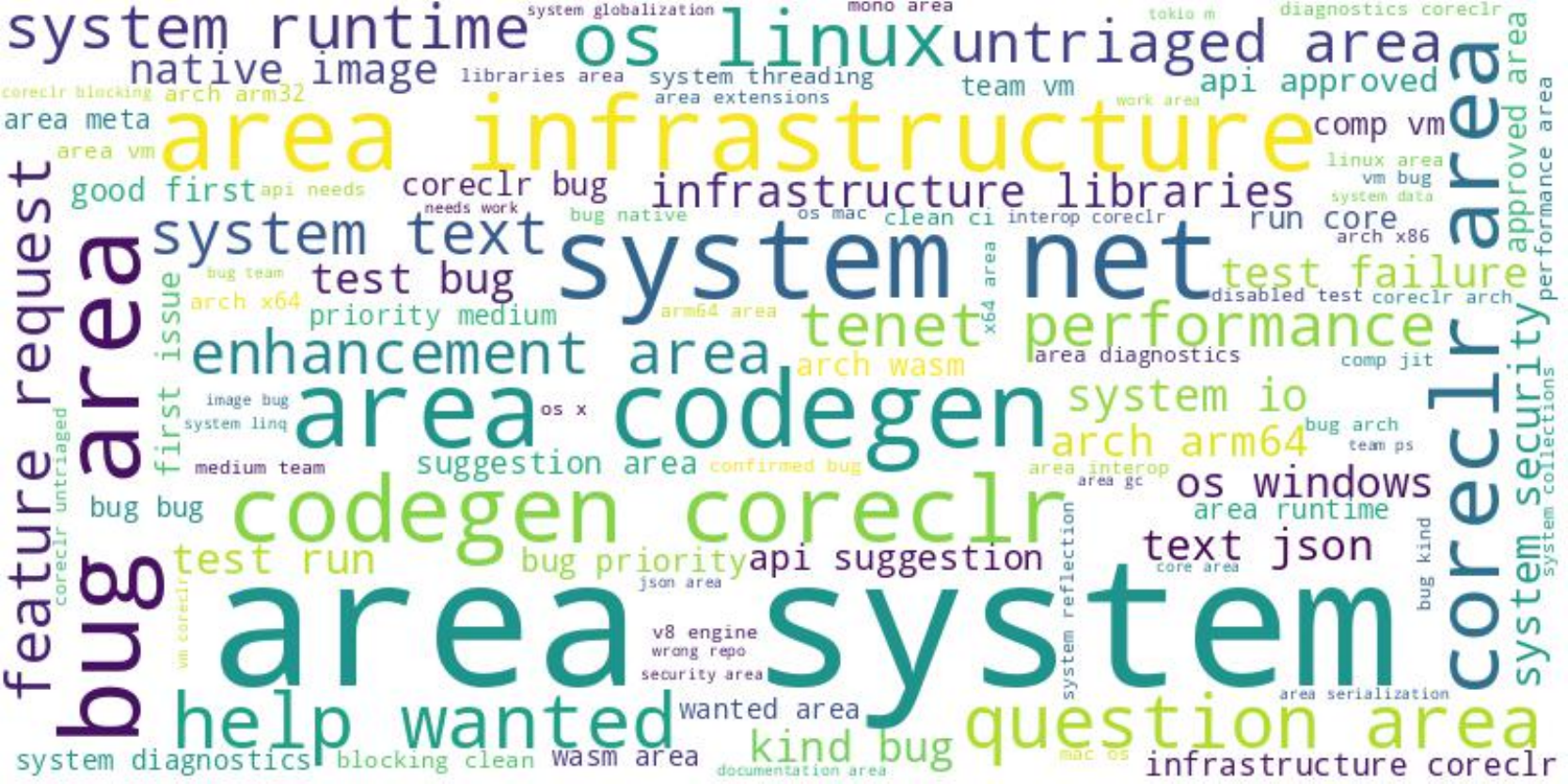}
\caption{Word cloud of closed issue labels}
\label{fig:wordcloud}
\end{figure}

We also investigated issues and repositories that utilized the assignee feature for issue management and found that 85.29\%(29) repositories use this feature and only a few issues, 28.30\% (33336) of all have assignees. Among the closed issues, 30.30\% (29515) and among the open issues, 18.74\% (3821) are assigned to the developers. We further investigated the presence of issue comments whether for issue management and communication, discussion forum is being used or not. Our result shows that 90.65\% (106772) issues of all contain at least one comment. Among the closed and open issues,  92.34\% (89940) and 82.57\% (16832) issue contain comments. The average comment count is 5.89 with a standard deviation of 9.69 and median is 3 comments per issue and maximum is 878 comments.

\begin{tcolorbox}[flushleft upper, boxrule=1pt, arc=0pt, left=0pt, right=0pt, top=0pt, bottom=0pt, colback=white, after=\ignorespacesafterend\par\noindent]
\textbf{Summary of RQ3.} In this section, we present the result of our empirical study to show how issues are managed using labels and assignee in the runtime system repositories. Though most of the repositories use these features, on issue level,  82.65\% issues have label while only  28.30\% (33336) of all issues have assignees. Also,  90.65\% issues at least one comment. 
\end{tcolorbox}

\subsection{RQ4. What are the relationships between metadata used to describe the issues and the closure of issues?}\label{sec:relationship}


\subsubsection{Approach} We selected total 11 metadata from issues and repositories and  hypothesised these features can possibly affect the issue closure. As a first feature, we selected the presence of labels on issues and  measured its effect on issue closure. To ease issue categorization, GitHub comes up with a feature for tagging the issues with related labels. Though GitHub offers nine default labels to start with: bug, documentation, duplicate, enhancement, good first issue, invalid, question and wontfix;  the repository contributors with triage access can create their own customized and project specific labels for tagging their issues. A label helps the stakeholders to easily understand what type of issue this is and also the use of labelling help maintainers managing issues. If the label is not added from the poster and maintainers tagged it; then having labels on this issue can indicate that this issue has been noticed and more likely to be assigned and addressed soon. Also, issues can be tagged with multiple labels to categorize it in a specific topic. For example, 'bug' label is typically used for any bug related issue but it does not define in which module this bug occurs. In that case, labelling with module tag may help stakeholders to understand exactly where the bug is. We have also checked if the number of labels have any effect on issue closure.  

Next, we investigated another GitHub Feature regarding issue assignment. Whenever an issue is noticed, an expected practice is to assign issues to the experts on that field. An assigned issue is more likely to address faster as the assigned person is more likely to start working on it as soon as possible. However, as we saw in \sec\ref{sec:issuemanagement},only around 30\% issues are assigned; that is because in many cases, developers are called for resolution in the comment section or by direct communication. Issues can be assigned to multiple developers so that they can join their hands on resolving and help each other.  We'll examine if the presence of assignee or number of assigns have any relationship with the closure of the issue. 

Next we looked at issue description to get several metadata. For example, we checked existence of code in issue body and sentiment of issue title and body. Some issues contain code blocks such as: snippets of the portion of the code that results in wrong output. Code blocks can help managers or developers to understand the issue more clearly. A clear understanding of an issue is a first step for addressing an issue. Again, The sentiment of a title and body ( positive, negative and neutral) are investigated using a machine learning sentiment classifier and checked if any kind of negativity expressed in the issue introduction make maintainers reluctant to start working on it.  We used BERT4SentiSD \cite{senti4bert} to detect the sentiment of our desired texts. Length of issue title and body are also checked. As title is supposed to convey the main point of the issue in one line and make maintainers notice, it can be assumed the longer the title is, the more information it carries.  Again the more informative the title  is, the easier it will be to understand at a glance, even before going through the issue description which will then turn to address the issue. Similar hypothesis is also built for issue body. From the issue title and body, we obtain the length of these strings using \textit{len()} function. We also investigated the issue discussion;  since when an issue is noticed and developers start addressing they might put comments on discussion forum. Discussing among themselves can help them understand the issue clearly and come up with a solution quickly. At the end, we made the assumption that a repository is more likely to address more bugs the more contributors it has. Therefore, we investigated the relationship between number of contributors and issue closure. 

For measuring effects, we grouped the issues on two categories based on the status of the issues: open issues and closed issues and used Cohen’s effect size $\delta$ \ to calculate the strength of the relationship between the features described above of open and closed issues. For interpreting the effect size, we used the commonly used guidelines for interpretation which suggests based on the benchmark that effect size will be regarded as negligible (N) if the value less than 0.2; small (S) if value is between 0.2 and 0.5; medium (M) if value between 0.5 to 0.8 and large (L) if it is greater than 0.8  We also used Wilcoxon rank-sum test \cite{Wilcoxon1992} to measure the significance of the differences between features of open and closed issues.

\subsubsection{Result} \tbl\ref{table:relationship} presents the relationship of 11 features with issue closure by showing the values of the Cohen’s $\delta$ and effect size \ along with the statistical significance from Wilcoxon rank-sum test \cite{Wilcoxon1992}. This result shows that all the features show a statistically significant difference between the open and closed issues except sentiment of title and body. Again, the effect size of has-label is small (0.23); meaning labelled issues are more likely to be closed. Similarly, has-assignee, num-assignee and has-comment has a small effect size of 0.27, 0.24  and 0.3 respectively which indicates that issues are more likely to be closed if these have assignees or there are multiple assignees or there are discussion under the issues. Also, num-contributors feature also shows a small effect size of 0.24 indicating  the more contributors in a repository, the more issue will be closed. The other remaining features shows negligible effect size.


\begin{table}[t]
\rowcolors{2}{}{lightgray!30}
    \centering
        \caption{The relationship of different metadata with issue closure (E = Effect size)}
    \label{table:relationship}
    \begin{tabular}{l|l|c|c}
    \toprule
        Feature &  Description & $\delta$ & E \\ \midrule
        has-label & Presence of labels in an issue & 0.23*** & S\\
        num-label & Number of labels in an issue & 0.17*** & N\\
        has-assignee & Whether the issue is assigned or not & 0.27*** & S\\
        num-assignee & Number of assignees for an issue & 0.24*** & S\\
        has-comment & Presence of issue discussion & 0.30*** & S\\
        has-code & Presence of code in issue body & 0.04*** & N\\
        sent-title & Negative sentiment on an issue title & 0.02 & N\\
        sent-body & Negative sentiment on the issue body & 0.00 & N\\
        title-length & Length of title string & 0.04*** &  N\\
        body-length &  Length of the issue body string & 0.05*** &  N\\
        num-contrib & Number of contributors in repository & 0.24*** &  S\\
        \bottomrule
    \end{tabular}
\end{table}

\begin{tcolorbox}[flushleft upper, boxrule=1pt, arc=0pt, left=0pt, right=0pt, top=0pt, bottom=0pt, colback=white, after=\ignorespacesafterend\par\noindent]
\textbf{Summary of RQ4.} In this section, we investigated whether there is any relationship between some features of issue and repository with issue closure. We found that presence of labels, assignees, comments and number of assignees and contributors of a repository can have a small effect on issue closure with statistically significant differences.
\end{tcolorbox}

\subsection{RQ5. Is there any relationship between the types of information discussed in issue comments and the closure of the issues?}\label{sec:infotype}


\subsubsection{Approach} We investigated the presence of different types on information in runtime system issue discussion and measured its effect on issue closure. Since our automated issue discussion type detector works with sentences; we extracted the sentences from the issue title, body and comments and obtained total 10111 sentences from sample issues. Then we replaced the code snippets with CODE keyword, reference links to external resources with URL keyword and trimmed the sentences.  Along with that, we also retrieved metadata of the sentences following the guidelines of Arya et al. \cite{InfoTypesOSSDiscussion}. These metadata include  participant information, sentence length related information, structural (position) and temporal (time) information and existence of code snippets in a sentence. 

The retrieved features are then fed to a machine learning model of lightGBM for information type detection. To train this model, we have used Arya et al.'s \cite{InfoTypesOSSDiscussion} curated dataset which they created by manually labelling the sentences of 15 issues of three AI GitHub repository. Though initially their dataset contains 16 information types, some labels do not contain necessary number of data and some have multiple labels. We discarded those from the dataset and used the final one containing 4330 sentences for training. Our model gets 64\% f1-score with 10-fold cross validation and can detect following 13 issue discussion information types: Action on Issue, Bug Reproduction, Contribution and
Commitment, Expected Behaviour, Investigation and Exploration,  Potential New Issues and Re quests,Motivation, Observed Bug Behaviour, Social Conversation, Solution Discussion, Solution Usage,Task Progress and Workarounds. We measured the distribution on information types in these issue discussion. Next for measuring effect of these types on issue closure, we followed the previous techniques of RQ4; however here we discarded the sentences of issue title or body and only consider issue comments.

\subsubsection{Result}  \fig\ref{fig:infotypes} presents the distribution of information types present in the retrieved sentences. Solution discussion is the mostly found information in runtime system issues covering almost half (48.66\%) of the sentences ; followed by social conversation covering around one-fourth (25.98\%) of all sentences. Potential New Issues and requests, Observed Bug Behaviour, Motivation and Action on issue contains 7.75\%, 3.61\%, 3.56\%
and 3.47\% of all sentences. Others combined contain only 6.96\% of the sentences.
\begin{figure}[t]
    \centering
\begin{tikzpicture}  
\begin{axis}  
[   axis x line*=bottom,
     axis y line*=left,
     height = 60mm,
    xbar,  
    xlabel={\ Sentences}, 
    bar width=4pt,
    symbolic y coords={ Expected Behaviour, Workarounds, Contribution and Commitment, Solution Usage,Bug Reproduction,Task Progress,Investigation and Exploration, Action on Issue,  Observed Bug Behaviour, Motivation, Potential New Issues.,Social Conversation, Solution Discussion},
    ytick=data,  
    nodes near coords, 
    nodes near coords align={horizontal},  
    ]  
\addplot [fill=gray,
        ]coordinates {(358,Action on Issue) (119,Bug Reproduction) (50,Contribution and Commitment) (25,Expected Behaviour) (227,Investigation and Exploration)
(365,Motivation) (364,Observed Bug Behaviour) (784,Potential New Issues.) (2619,Social Conversation) (4916,Solution Discussion) (190,Task Progress) (39,Workarounds) (55,Solution Usage)
};  
\end{axis} 
\end{tikzpicture}  
\caption{Info type distribution in sample issues' discussion}
\label{fig:infotypes}
\end{figure}
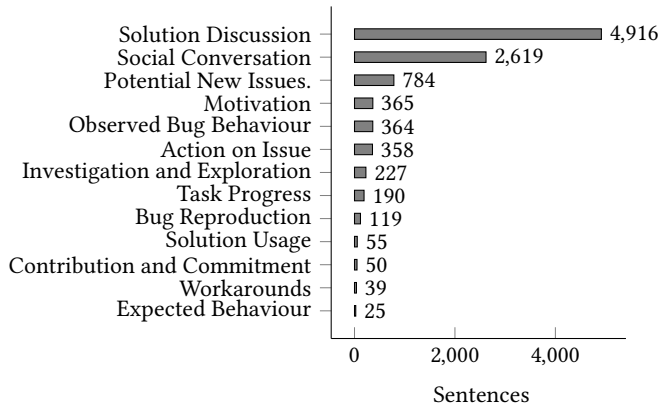

\begin{table*}[t]
\rowcolors{2}{}{lightgray!30}
    \centering
    \caption{The relationship of different information type found in issue comments with issue closure (E = Effect size)}
    \label{table:infotypeeffect}
    \begin{tabular}{l|l|c|c}

    \toprule
        Information Type & Description & $\delta$ & E \\ \midrule
        Observed Bug Behaviour & Any question or report on observed bug behaviour & 0.24 & S \\ 
        Investigation and Exploration & Suggestion, statement or report on exploring the cause of the issue & 0.23 & S \\ 
        Expected Behaviour & Describing ideal scenario from user/personal feedback & 0.18 & N \\ 
        Solution Discussion & Suggestion, statement or report on solution implementation and challenges & 0.16 & N \\ 
        Motivation & Reason for issue resolution or enhancement implementation & 0.16 & N \\ 
        Bug Reproduction & Question or report on bug reproduction & 0.14 & N \\ 
        Task Progress & Asking or providing task update & 0.11 & N \\ 
        Social Conversation & Small talk, showing gratitude or other emotions &0.09 & N \\ 
        Contribution and Commitment & Comments regarding willing to contribute or assigning for contribution & 0.05 & N \\ 
        Usage & Question or suggestion on new solution usage &0.04 & N \\ 
        Action on Issue & Issue management actions like close, labelling, reopening etc. & 0.03 & N \\ 
        Workarounds & Sharing temporary solution & 0.01 & N \\ 
        Potential New Issues and Requests & Identifying new bugs or requesting modification while investigating an issue &0.01 & N \\ \bottomrule
    \end{tabular}
\end{table*}

Next, we measured the effect of the presence of certain information types in issue comments with issue closure. From the analysis we could not find any significantly significant relations. \tbl\ref{table:infotypeeffect} presents the result of this analysis. We also investigated the sample issues to check whether there is any relationship between issue discussion sentiment with issue closure rate. Here also, it does not show any statistically significant difference between the open and closed issues.

\begin{tcolorbox}[flushleft upper, boxrule=1pt, arc=0pt, left=0pt, right=0pt, top=0pt, bottom=0pt, colback=white, after=\ignorespacesafterend\par\noindent]
\textbf{Summary of RQ5.} We identified 13 information types in the issue report comments. Solution Discussion is the most found information type among the 13 information types. No information type has statistically significant correlation with issue closure.
\end{tcolorbox}

In this section, we share some recommendations to runtime system project maintainers for effective management along with the threats to validity of this work.

\subsection{Recommendation} \label{sec:reco}
The result of our empirical study shows issues are more likely to be closed when they have assignees. However currently only around 30\% issues have designated assignees. We recommend runtime system repository maintainers to use the assignee feature provided by GitHub for effective issue management. There are different tools that can be used for automated triaging \cite{AutomatedBugTriagingTopicModel,triagetoss,ImprovingBT,Li2021RevisitingTF, Zhang2017EnLDAAN,Kardoost2022DevRankerAE, Tao2017BugRE,Xi2018AnEA,Alazzam2020AutomaticBT}  and reduce manual work; maintainers can use these for this task. 
\begin{Recommendation}
 Runtime system project maintainers should assign issues to responsible experts as soon as an issue is reported. 
\end{Recommendation}

Again, we observed that having labels on the issues also affect the issue closure. Kim et al. \cite{multilabel} showed that projects that use multi-label features can manage issues more effectively. 
Therefore, it is recommended to runtime system repository stakeholders to actively leverage this feature. They can use some automated issue labelling tools \cite{predictissuetype,tickettagger} for speeding up their work.

\begin{Recommendation}
 Runtime system project maintainers should tag issues with labels for effective issue management.
\end{Recommendation}

Apart from these, issue discussions also help an issue closure. Developers can be more clear of the issue from the reporter, discuss with their peers about issue investigation and  solution and can come up with a solution. Our result also shows, half of the discussion consists of only Solution Discussion. Therefore, we encouraged developers to actively engage on discussion if they find any trouble on understanding any issue and need help from peers and not hesitate or fear to ask.

\begin{Recommendation}
Developers should engage in discussion to be clear of the issues and discuss idea about investigation and solution. 
\end{Recommendation}

Again, descriptive and positive issue description can create a significant difference to the closure of the issue. We suggest issue reporters to describe the issues with no negative expression; explain all the details of the issue elaborately and provide all sorts of information. 

\begin{Recommendation}
Issue reporters should elaborately explain the issues. 
\end{Recommendation}

Also, our result suggests presence of code can cause a significant difference to issue closure. Therefore, we would like to suggest issue reporters to add code blocks in markdown syntax for issues like bug, error or wrong output when available. This way, issue will be more precise and easy to understand for maintainers and developers.

\begin{Recommendation}
Issue reporters should add code snippets if available for describing an issue like bug, error or wrong output. 
\end{Recommendation}

From our analysis, we found that besides error, users often ask questions and request for enhancement. To address and reduce these issues, it is recommended to provide more clear instructions on how to use each function and strengthen documentations ; also share future plan of enhancement if possible. 

\begin{Recommendation}
Project maintainers should strengthen documentations and provide clear instructions of usage.
\end{Recommendation}

\subsection{Threats to Validity}\label{subsec:validity}
\textbf{Internal validity} threats refer to the authors' biases while conducting a study. In order to mitigate the bias for manual labelling, we picked issues randomly for creating the sample dataset and picked at least one comment from each repository. By this way, we ensured that we do not select only certain types of issues and we do not discard any repositories.  \textbf{Construct validity} threats refer to the error in data collection process.  To identify the runtime system projects, we used both topics and the project description metadata and followed best practices to analyze GitHub repositories. \textbf{External validity} threats concern the generalizability of the findings of this study. Since we are focusing on runtime system issues, we tried to gather as much as GitHub hosted runtime system projects and finally utilised 34 projects. We considered all issues for issue management analysis. For issue content analysis, we took a statistically significant sample. Our model can be applied to any similar platforms to get the empirical result about the issue addressing and management along with retrieving the distribution of information types in issue discussion.
\section{Related Works} \label{sec:related}
Related work can broadly be divided into two categories:  \begin{inparaenum}
\item Mining GitHub issues and
\item Empirical study on GitHub community
\end{inparaenum}. 
\subsection{Empirical study on GitHub community}
There are numerous studies in literature where researchers empirically investigate different communities as per their interest to identify
unique properties, pattern and trend. These studies provide with necessary takeaways and recommendation that would be useful to that community. Gonzalez et al. \cite{stateofai} studied 10 years of history of Artificial Intelligence (AI) and Machine Learning (ML) GitHub community, found unique patters and current states and measured the collaboration and autonomy of the developers within repositories. Similarly, Das et al. \cite{blockchain} focused on blockchain repositories. Yang et al. \cite{AIrepo} again investigated AI community to find out issue reporting, addressing and management pattern while Wang et al. \cite{covid} aimed at COVID-19 themed repositories. However, to the best of our knowledge, no empirical study has been done on runtime system repositories even though people are engaged in this community for years. To fill up this gap and support the interest of our industrial partner X, we consider runtime system repositories for our empirical study.    

\subsection{Mining GitHub Issues}
Researchers investigated bugs or specific kind of issues on Open Source Software (OSS) GitHub repositories. Chen et al. \cite{dormantbug} examined dormant bugs in OSS repositories, Makkouk et al. \cite{perfbug} studied performance bugs in deep learning frameworks and Zahedi et al. investigated security issues \cite{securityissue} of OSS projects. Our work focuses on issue reports in runtime system repositories and analyze them for categorization and create a issue type taxonomy. 

Along with issue reports, issue discussions are also studied in literature to find sentiments and information types associated with them. Arya et al. \cite{InfoTypesOSSDiscussion} uncovered 16 information types from issue discussion and developed shallow ML models for information type detection where  Mehder and Aydemir  \cite{Mehder2022ClassificationOI} used deep ML models. Our study used Arya et al.'s \cite{InfoTypesOSSDiscussion} taxonomy to find information type distribution in runtime system repositories. On the other hand, Guzman et al. analyzed \cite{SentimentAnalysisCommitComments} sentiment of commit comments where Murgia et al. \cite{EmotionInIssueCommentOSS} analyzed issue comments. There are other work to find negativity in comments such as: track down uncivil \cite{IncivilityInOSS} and toxic \cite{Didyoumissmycomment} comments.  

Besides analyzing issue contents, several works focus on investigating issue addressing and management in GitHub repositories. Kikas et al. \cite{issuedynamic} studied the temporal dynamics of issues of 4000 GitHub projects and analyzed how issue creation rate, number of pending issues and average lifetime evolve over the time. Kim et. al. \cite{multilabel} conducted an empirical study on usage of multiple labels. Mumtaz et al.\cite{issuecommenter} investigated the impact of one GitHub features,  “assign issues to issue commenters” on the social structure of the projects. Our empirical study look at the runtime system issue management states, pattern and trend such as: issue close rate, ignore rate, addressing time and find the relationship of different metadata and GitHub features with issue closure.

\subsubsection{Result} \tbl\ref{table:relationship} presents the relationship of 11 features with issue closure by showing the values of the Cohen’s $\delta$ and effect size \ along with the statistical significance from Wilcoxon rank-sum test \cite{Wilcoxon1992}. This result shows that all the features show a statistically significant difference between the open and closed issues except sentiment of title and body. Again, the effect size of has-label is small (0.23); meaning labelled issues are more likely to be closed. Similarly, has-assignee, num-assignee and has-comment has a small effect size of 0.27, 0.24  and 0.3 respectively which indicates that issues are more likely to be closed if these have assignees or there are multiple assignees or there are discussion under the issues. Also, num-contributors feature also shows a small effect size of 0.24 indicating  the more contributors in a repository, the more issue will be closed. The other remaining features shows negligible effect size. \section{Conclusions} \label{sec:conclusion}
This paper contributes to understand issue content and management of runtime system repositories by conducting an empirical study of 111789 issues of 34 projects. First, we analyzed the issue contents to determine the types of the reported issues. We found 10 types of issues are logged in the runtime system repos, such as Enhancement, Test Failure and Bug, etc. We next analyzed how the issues are managed. We found out a very small number of issues (0.69) are ignored on these repositories and 82.69\% issues are resolved with a median issue close rate of 76.1\% and median addressing time of 58 days. Our findings also show that though labelling and issue comments are used for issue management; only around 30\% issues have designated assignees. However, our result shows having assignee can affect issue closure.  Also, on sample issues, Solution Discussion and Social Conversation are found to be the most commented topic on issue discussion. Our future work will focus on the development of processes, techniques to support the efficient management and fixing of runtime system issues. We also intend to facilitate efficient collaboration among the stakeholders based on context-aware software and issue analytics. 

\section*{Data Availability}
The code and data used for this paper are shared in this anonymous repo: \url{https://anonymous.4open.science/r/EmpiricalRuntime-25EB}

\begin{small}
\bibliographystyle{abbrv}
\bibliography{References}
\end{small}
\end{document}